\documentstyle[aps,prb]{revtex}
\begin{document}
\draft
\title{Hall potentiometer in the ballistic regime}
\author{B. J. Baelus and F. M. Peeters\cite{refpeeters}}
\address{Departement Natuurkunde, Universiteit Antwerpen (UIA),\\
Universiteitsplein 1, B-2610 Antwerpen, Belgium}
\date{\today }
\maketitle

\begin{abstract}
We demonstrate theoretically how a two-dimensional electron gas can be used
to probe local potential profiles using the Hall effect. For small magnetic
fields, the Hall resistance is inversely proportional to the average
potential profile in the Hall cross and is independent of the shape and the
position of this profile in the junction. The bend resistance, on the other
hand, is much more sensitive on the exact details of the local potential
profile in the cross junction.
\end{abstract}

\pacs{73.40.-c; 03.65.Sq; 73.50.Jt}

A conductive atomic force microscope\ (AFM) tip has been used, in contact 
\cite{trenkler} and non contact \cite{cunningham} mode,\ as a local voltage
probe in order to measure the distribution of the electrical potential on a
surface. This technique is called {\it nanopotentiometry} and allows
two-dimensional potential mapping \cite{trenkler,cunningham}. The
sensitivity and the spatial resolution are limited by the finite size of the
conductive probe and by the quality of the surface preparation. On the other
hand such tips can also be used to induce potential variations in the sample
in order to influence the conduction locally \cite{eriksson}. Measuring the
change in the resistance of the device gives information on the local
transport properties. Due to the complexity of the problem, i.e. the
different dielectric layers, interfaces and screening of the two-dimensional
electron gas (2DEG), it is rather difficult to calculate theoretically the
induced potential of the tip in the 2DEG. The aim of the present work is to
propose a new technique which allows to measure such local potential
profiles inside a 2DEG. Using the Hall effect we will demonstrate
theoretically that the 2DEG can be used as a probe to measure the
inhomogeneous induced potential profile. The system we consider is depicted
in the inset of Fig. 1(a) and consists of a mesoscopic Hall bar placed in a
homogeneous magnetic field, containing a local inhomogeneous potential
profile in the cross junction which is e.g. induced by a STM-tip.

To describe the transport properties in the Hall cross we will use the
billiard model \cite{beenakker}. In this model the electrons are considered
as point particles which is justified when the Fermi wavelength $\lambda
_{F}\ll W,d$ where $2W$ is the width of the Hall probes and $d$ the radius
of the potential profile which acts as a scatterer for the electrons. The
electron motion is taken ballistic and governed by Newtons law which is
justified at low temperatures and in case of high mobility samples where the
mean free path $l_{e}\gg W,d$. We assume that the temperature is not
extremely low such that interference effects are averaged out due to thermal
smearing. In a typical GaAs-heterostructure the electron density is $n\sim
3\times 10^{11}\,cm^{-2}$ with typical low temperature mobility $\mu \sim
10^{6}\,cm^{2}/Vs$, which gives $\lambda _{F}=450\,$\AA\ and $l_{e}=5.4\,\mu
m$. This billiard model has been used successfully \cite{beenakker} to
describe e.g. the experiments of Ref.~\cite{ford} and to explain \cite{li}
the working of a ballistic magnetometer \cite{geim}.

Using the Landauer-B\"{u}ttiker formalism \cite{buttiker} for the Hall geometry with
identical leads, the Hall $R_{H}$ and the bend $R_{B}$\ resistances are
given by 
\begin{eqnarray}
R_{H} &=&\frac{\mu _{4}-\mu _{2}}{eI_{1\rightarrow 3}}=\frac{h}{2e^{2}}\frac{%
T_{R}^{2}-T_{L}^{2}}{Z},  \label{eq1} \\
R_{B} &=&\frac{\mu _{2}-\mu _{3}}{eI_{1\rightarrow 4}}=\frac{h}{2e^{2}}\frac{%
T_{F}^{2}-T_{L}T_{R}}{Z},  \label{eq2}
\end{eqnarray}
where $Z=\left[ T_{R}^{2}+T_{L}^{2}+2T_{F}\left( T_{R}+T_{F}+T_{L}\right) %
\right] \left( T_{R}+T_{L}\right) $ and $T_{R}$, $T_{L}$, $T_{F}$ are the
probabilities for an electron to turn to the right probe, to the left probe
and to the forward probe, respectively. These probabilities will be
calculated using the ballistic billiard model \cite{beenakker,li}.

In the following we will express the magnetic field in units of $%
B_{0}=mv_{F}/2eW$, and the resistance in $R_{0}=\left( h/2e^{2}\right) \pi
/2k_{F}W$, where $W$ is the half width of the leads, $m$ is the mass of the
electron, $k_{F}=\sqrt{2mE_{F}/\text{%
%TCIMACRO{\UNICODE{0x127}}%
%BeginExpansion
h\hskip-.2em\llap{\protect\rule[1.1ex]{.325em}{.1ex}}\hskip.2em%
%EndExpansion
}^{2}}$, and $v_{F}=$%
%TCIMACRO{\UNICODE{0x127}}%
%BeginExpansion
h\hskip-.2em\llap{\protect\rule[1.1ex]{.325em}{.1ex}}\hskip.2em%
%EndExpansion
$k_{F}/m$ the Fermi velocity. For electrons moving in GaAs ($m=0.067\,m_{e}$%
) and for a typical channel width of $2W=1\,\mu m$ and a Fermi energy of $%
E_{F}=10\,meV$ $\left( n_{e}=2.8\times 10^{11}\,cm^{-2}\right) $, we obtain $%
B_{0}=0.087\,T$ and $R_{0}=0.308\,k\Omega $.

In order to demonstrate the main physics involved, we consider first a
mathematically simple potential profile, namely a rectangular potential
barrier with radius $d$ and height $V_{0}$ placed in the center of the cross
junction: $V\left( r\right) =V_{0}$ if $r<d$ and $V\left( r\right) =0$ if $%
r>d$. This potential barrier is schematically shown in the inset of
Fig.~1(a) by the shaded circular area. Inside the potential barrier the
kinetic energy of the electrons is reduced to $E_{F}-V_{0}$ with $E_{F}$ the
kinetic energy of the electrons outside this region. Hence, the electron
velocity $v$, the density $n$ of the electrons and also the radius $R_{c}$
of the cyclotron orbit are reduced inside the potential region.
For $V_0 < 0$ the opposite occurs. This will
result in changes in the transmission probabilities $T_{R}$, $T_{F}$, $T_{L}$
and consequently it will alter the Hall and bend resistances.

In Fig.~1 we show the Hall resistance $R_{H}$ and the bend resistance $R_{B}$
as function of the external applied magnetic field for different sizes and
heights of the rectangular potential profiles. Fig.~1(a) shows the Hall
resistance and Fig.~1(b) the bend resistance for different potential heights 
$V_{0}$ but fixed radius $\hspace{0in}d=0.5W$, while in Fig.~1(c) and 1(d)
the radius $d$ is varied and the potential height $V_{0}=0.2E_{F}$ is kept
fixed. Notice that there exists a critical magnetic field $B_{c}=B_{c}(d)$,
such that for $B>B_{c}$ no electrons are entering the area of the potential
barrier, because their cyclotron radius is so small that they skip along the
edge of the probe without 'feeling' the potential barrier. For $B>B_{c}$ the
diameter of the cyclotron orbit, $2R_{c}=2v_{F}/\omega _{c}$, is less than
the distance between the edge of the rectangular potential barrier and the
corner of the cross junction. Therefore, the electrons do not feel the
presence of the $V\neq 0$ region in the cross junction and the Hall and bend
resistances equal the classical $2D$ values: $R_{H}/R_{0}=2B/\pi B_{0}$ and $%
R_{B}=0$. A simple calculation of this critical field gives $B_{c}/B_{0}=4/(%
\sqrt{2}-d/W)$, which results into $B_{c}/B_{0}=4.4,7.8,18.7$ for $%
d/W=0.5,0.9,1.2$, respectively, and which agrees with the results of
Figs.~1(a,c). For $B<B_{c}$ the electron trajectories are only weakly bend
And consequently they sample the $V\neq 0$ region. The latter region increases
(decreases) the
turning probability $T_{R}$\ and hence reduces (enhances) 
$R_{B}$ and enhances (reduces) $R_{H}$
as is clearly observed in Fig.~1 for $V_0>0$ ($V_0<0$).
For low potential barriers, i.e. $|V_{0}|\ll
E_{F}$, we have $R_{B}\approx 0$ for $B/B_{0}>2$ which is substantially
below $B_{c}$. The reason is that almost no electrons finish in probe 2 and
3, i.e. $T_{L}\approx 0\approx T_{F}$, and consequently $R_{B}=0$. At the
magnetic field $B=2B_{0}$ the cyclotron diameter, $2R_{c}$, equals the probe
width, $2W$.\ For higher potential barriers some of the electrons are
deflected on the potential barrier into probe 2 and hence $R_{B}<0$. This is
clearly observed in Fig.~1(b).

We found that, even in the presence of an inhomogeneous potential profile,
the Hall resistance is linear for small magnetic fields (i.e. $B\ll B_{c}$).
The slope increases with the radius and with the height of the rectangular
potential barrier. We analyzed the Hall factor $\alpha =R_{H}/B$ for $B\ll
\,B_{c}$ in Fig.~2(a) as function of the radius $d$ for $V_{0}=0.2\,E_{F}$
and in Fig.~2(b) as function of the height $V_{0}$ for $d=0.4\,W$. Notice
that the Hall factor increases with $d$ and $V_{0}$. Increasing $d$ or $%
V_{0} $ results in an increase of the average potential $\left\langle
V\right\rangle $ or in a reduction of the average electron density $%
\left\langle n_{e}\right\rangle $ in the cross junction. For a rectangular
potential profile we find for the average electron density in the cross
junction $\left\langle n_{e}\right\rangle =n_{e}\left[ 1-\pi \left(
d/2W\right) ^{2}V_{0}/E_{F}\right] $ from which we define an effective Hall
coefficient $\alpha ^{\ast }=\left( R_{H}/B\right) \left\langle
n_{e}\right\rangle /n_{e}$ which is shown by the dashed curve in Fig.~2.
Notice that $\alpha ^{\ast }$ is almost independent (within 2\%) of $d$ for $%
d/W<1.0$ if $V_{0}$ is constant and practically independent of $V_{0}$ for $%
V_{0}/E_{F}<0.5$ if $d$ is constant. For $V_{0}/E_{F}>0.5$ the potential
profile is no longer a weak perturbation on the electron motion and
consequently the Hall resistance can no longer be described in terms of an
average electron density in the cross junction. The fact that $\alpha ^{\ast
}$ is practically independent of $V_{0}$ and $d$ indicates that for low
magnetic fields the Hall resistance is completely determined by the average
potential in the cross region, and is independent of the detailed potential
profile as long as $\left\langle V\right\rangle /E_{F}<0.5$. This is our
major conclusion of this work, which will be confirmed further.

The bend resistance $R_{B}$, on the other hand, is much more sensitive to
the exact form of the potential barrier. For example: the bend resistance
for a wide, but low barrier ($d=0.9\,W$, $V_{0}=0.2\,E_{F}$) does not equal
to the bend resistance for a narrow, but high barrier ($d=0.5\,W$, $%
V_{0}=0.648\,E_{F}$), even though the average potential in the junction
region is the same (see Figs. 1(b,d)).

Next, we investigate the effect of the functional form of the potential by
considering, as an example, a gaussian potential profile in the center of the
junction: $V_{g}\left( r\right) =V_{g,0}\exp \left( -r^{2}/d_{g}^{2}\right) $
with $V_{g,0}$ the height and $2d_{g}$ the width at half height. In Fig.~3
we compare the Hall factors $\alpha $ and $\alpha ^{\ast }$ of the
rectangular potential (solid curves) with height $V_{0}=0.2\,E_{F}$ and
radius $d$, and the gaussian potential (symbols) with width $d_{g}=d$ and
height $V_{g,0}$ as a function of the radius $d$. $V_{g,0}$ is varied with $d
$ such that the average potential inside the cross junction is the same for
the two potential profiles $\left( \left\langle V\right\rangle =\left\langle
V_{g}\right\rangle \right) $. The differences between the effective Hall
factor $\alpha ^{\ast }$\ in the two cases is negligeable and hence this
illustrates again that the Hall resistance is completely determined by the
average potential in the cross region, and is independent of the detailed
potential profile. In the inset of Fig.~3 we show, as an example, the
rectangular potential with $d=0.5\,W$ and $V_{0}=0.2\,E_{F}$ and the
corresponding gaussian potential with $d_{g}=0.5\,W$ and $V_{g,0}\simeq
0.202\,E_{F}$ which has the same $\left\langle V\right\rangle $.

Finally, we investigate the effect of displacing the rectangular potential
barrier away from the center of the junction. As an example, we consider a
rectangular potential barrier with height $V_{0}=0.2\,E_{F}$ and radius $%
d=0.5\,W$ which is displaced at different distances $\rho /W=0,\pm 0.1,\pm
0.2,\pm 0.3,\pm 0.4,\pm 0.5$ from the center of the cross junction in
different directions $\varphi =0,\pi /4,\pi /2,3\pi /4$ with regard to the
x-axis. Notice that in all these cases the entire potential barrier is
inside the cross junction and hence the average potential in the cross
junction is the same. Because the problem is no longer symmetric we are not
allowed to use the reduced Eqs. (\ref{eq1}) and (\ref{eq2}) but we are
forced to retain the original Landauer-B\"{u}ttiker formula \cite{buttiker}.
In Fig.~4 we show that the change in the effective Hall factor $\alpha
^{\ast }$ as function of the distance $\rho $ for the different directions $%
\varphi $ is less than 1\% for $d=0.5\,W$ and $V_{0}=0.2\,E_{F}$. Only when
the circular potential barrier is very close to one of the probes the
deviation becomes of the order of 1\%.\ This result illustrates that for low
magnetic fields the Hall resistance is completely determined by the average
potential in the cross region, and is independent of the detailed position
of the potential barrier in the cross junction.

In conclusion, we investigated the Hall and the bend resistances, in the
presence of an inhomogeneous potential profile in the junction of a
mesoscopic Hall bar. We found that in the low magnetic field regime the Hall
resistance is linear in the magnetic field and is determined by the average
potential in the cross junction, independent of the shape and the position
of the potential barrier as long as $\left\langle V\right\rangle /E_{F}<0.5$%
. This general result makes such a Hall device a powerful experimental tool
for non invasive investigations of induced potential profiles. The bend
resistance depends much more sensitively on the detailed shape and position
of the potential profile. The resistance at which the classical 2D Hall
resistance is recovered gives us information on the size of the potential
barrier. The present results are valid in the ballistic regime and are
expected, as in the case of the Hall magnetometer \cite{ibrahim}, to be
modified in the diffusive regime.

{\em Acknowledgments:} This work is supported by IUAP-IV, the Flemish
Science Foundation (FWO-Vl) and the Inter-University Micro-Electronics
Center (IMEC, Leuven).

\begin{figure}[tbp]
\caption{The Hall resistance $R_{H}$ (a) and the bend resistance $R_{B}$\
(b) as function of the magnetic field $B$ for different heights $V_{0}$ of
the rectangular potential barrier and for fixed radius $d=0.5W$. The Hall
resistance $R_{H}$ (c) and the bend resistance $R_{B}$\ (d) as function of
the magnetic field $B$ for different radii $d$ and for a fixed\ height $%
V_{0}=0.2E_{F}$.}
\label{fig1}
\end{figure}

\begin{figure}[tbp]
\caption{The Hall factor $\protect\alpha =R_{H}/B$ in the low magnetic
field\ region $\left( B=0.4B_{0}\ll B_{0}\right) $ as a function of (a) the
radius $d$ of the rectangular potential for $V_{0}=0.2E_{F}$, and (b) the
height $V_{0}$ of this potential for $d=0.4W$. The dashed curves are
obtained from the effective Hall factor $\protect\alpha ^{\ast }=\left(
R_{H}/B\right) \left\langle n_{e}\right\rangle /n_{e}$, where $\left\langle
n_{e}\right\rangle $ is the average electron density in the junction.}
\label{fig2}
\end{figure}

\begin{figure}[tbp]
\caption{The Hall factors $\protect\alpha $ en $\protect\alpha ^{\ast }$
resulting from a rectangular potential (solid curves) in the center of a
Hall cross are compared to these resulting from a gaussian potential
(symbols) which has the same average potential in the cross junction.\ In
the inset, the gaussian potential which corresponds to a rectangular
potential with $d=0.5W$ and $V_{0}=0.2E_{F}$ is shown.}
\label{fig3}
\end{figure}

\begin{figure}[tbp]
\caption{The effective Hall factor $\protect\alpha ^{\ast }$ for different
positions of the rectangular potential with $d=0.5W$ and $V_{0}=0.2E_{F}$
for $B=0.4B_{0}\ll B_{c}$. The potential barrier is displaced over a
distance $\protect\rho $ in a direction $\protect\varphi $ with respect to
the direction of the current (see inset). The curves are guides to the eye.
The effective Hall factor is scaled to its value for a rectangular potential
in the center of the Hall cross.}
\label{fig4}
\end{figure}

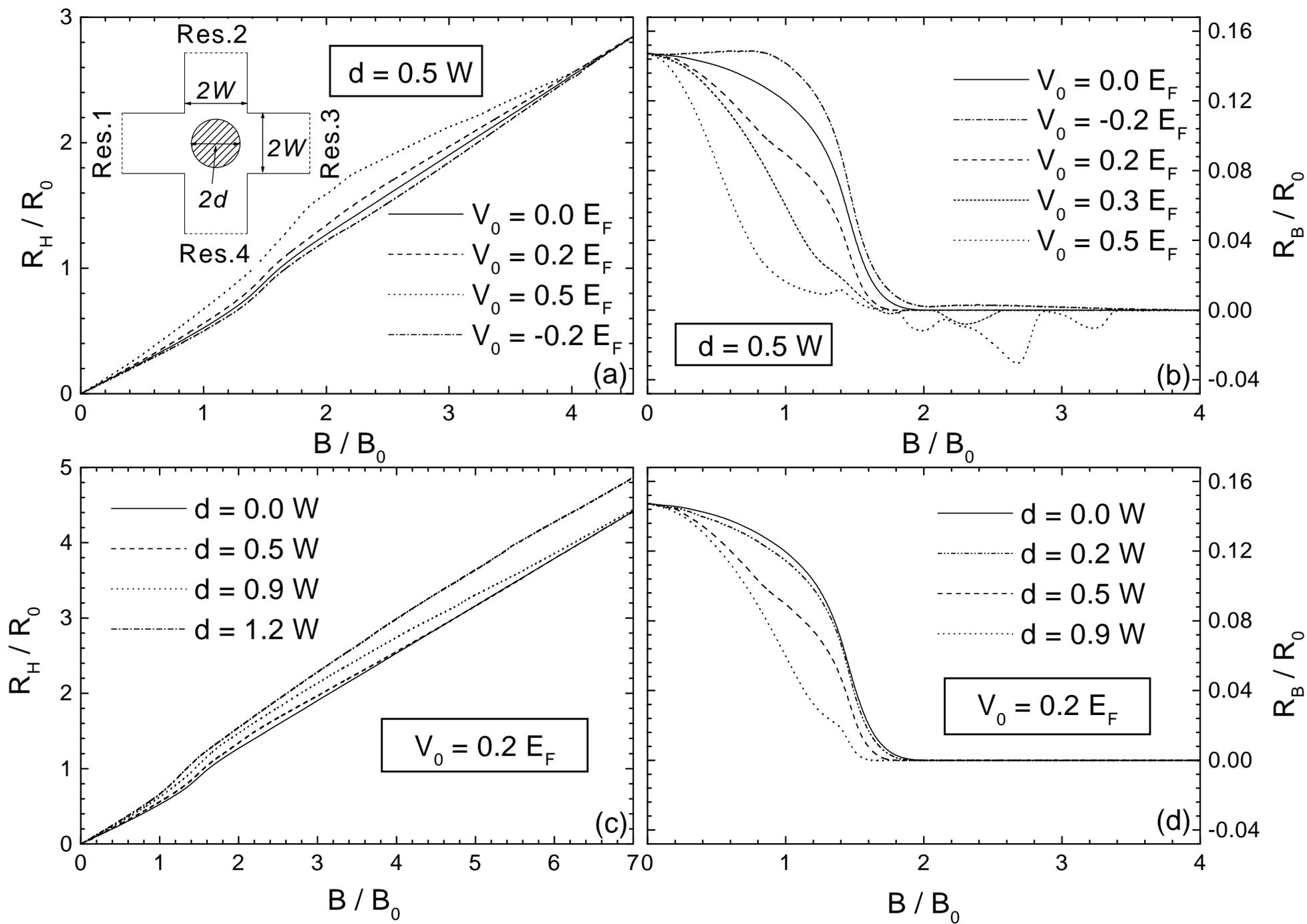
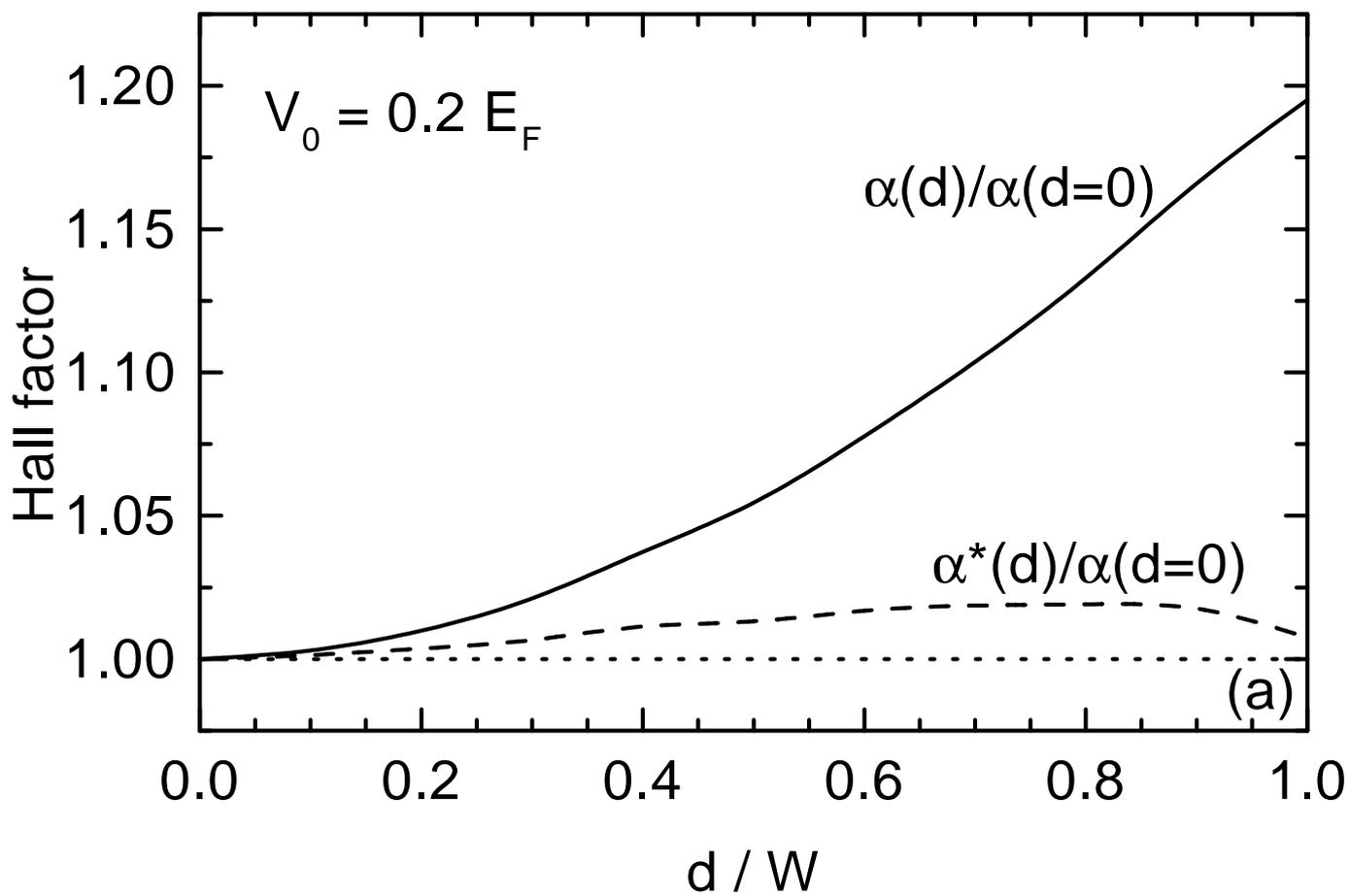
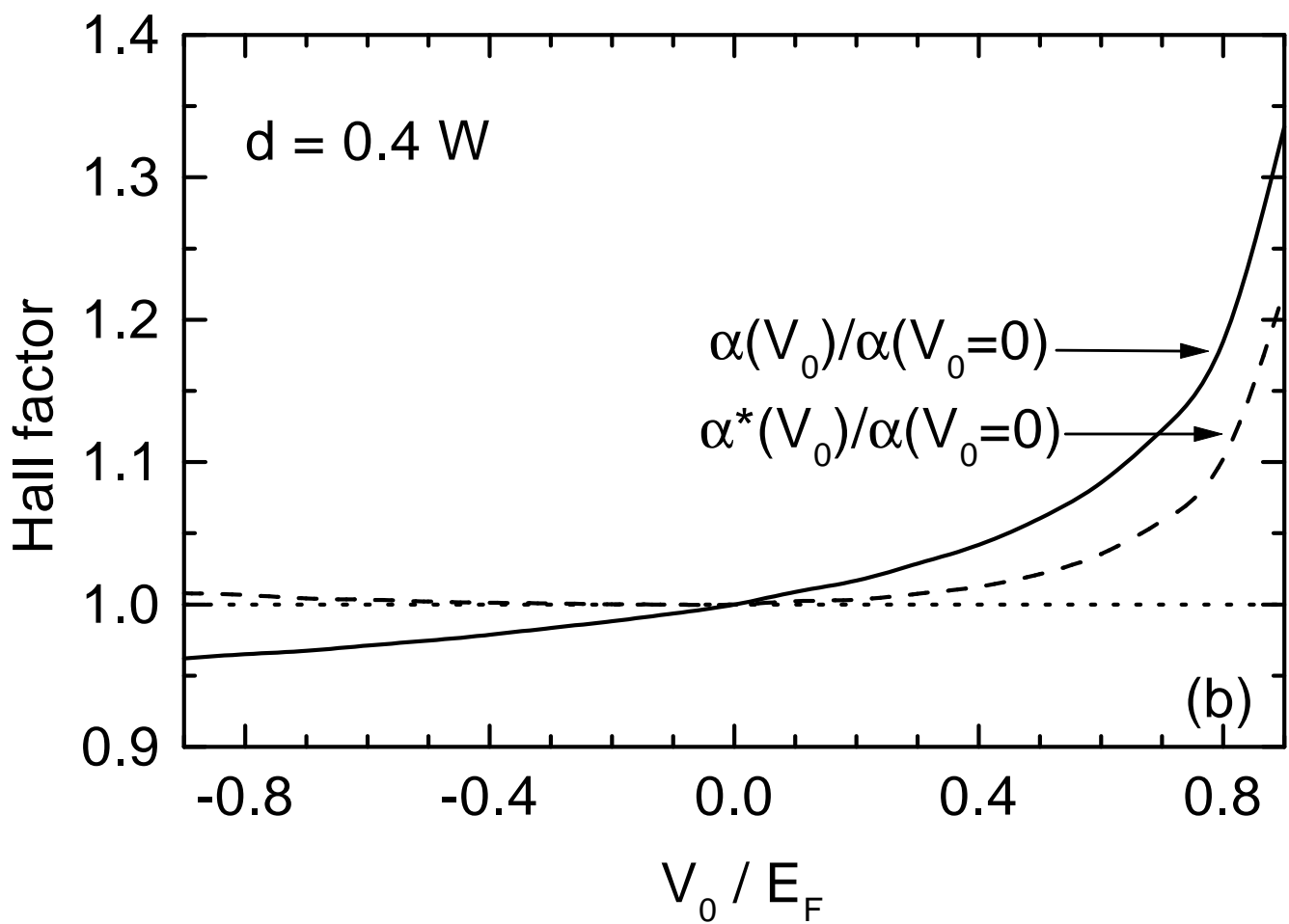
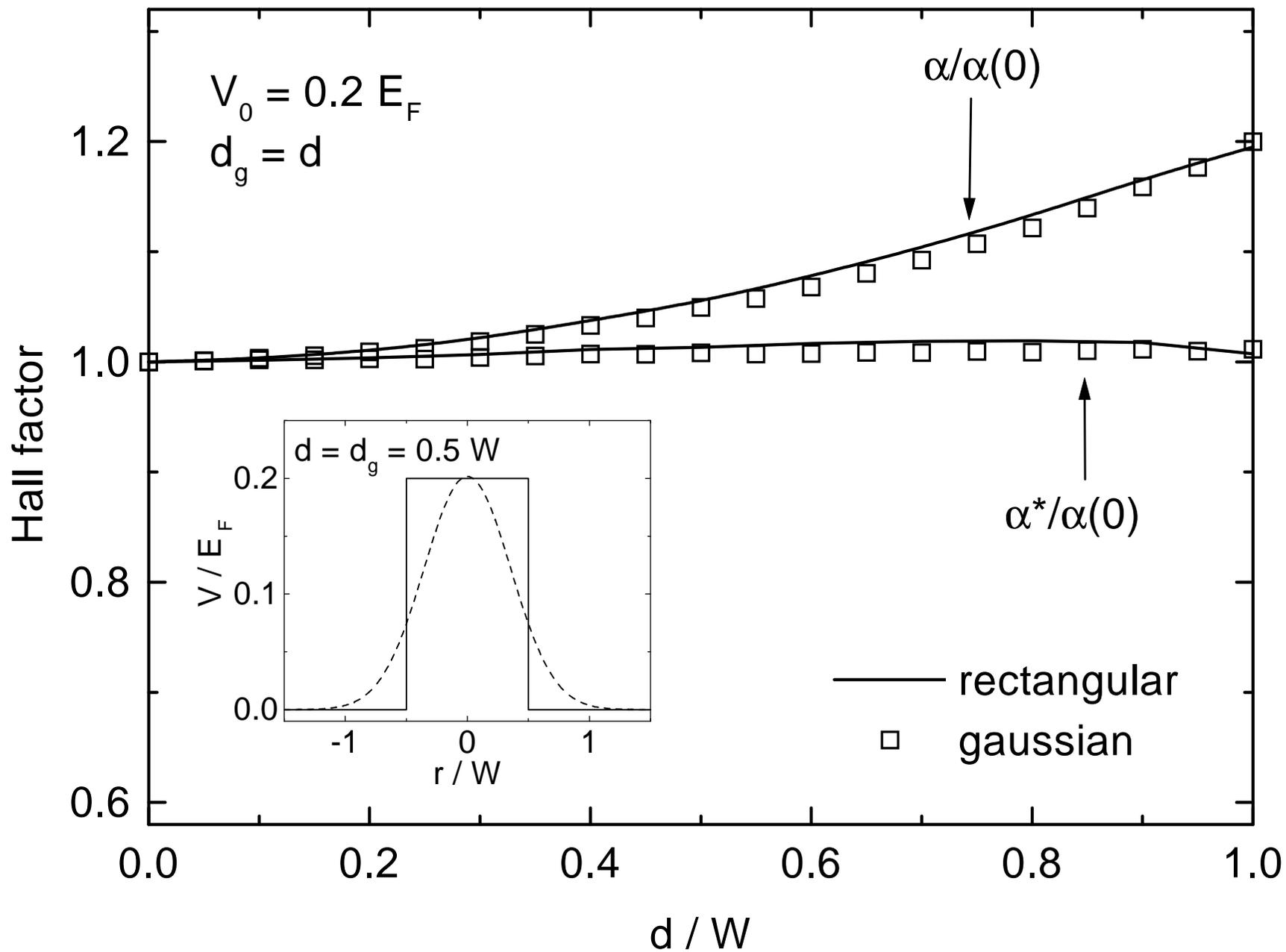
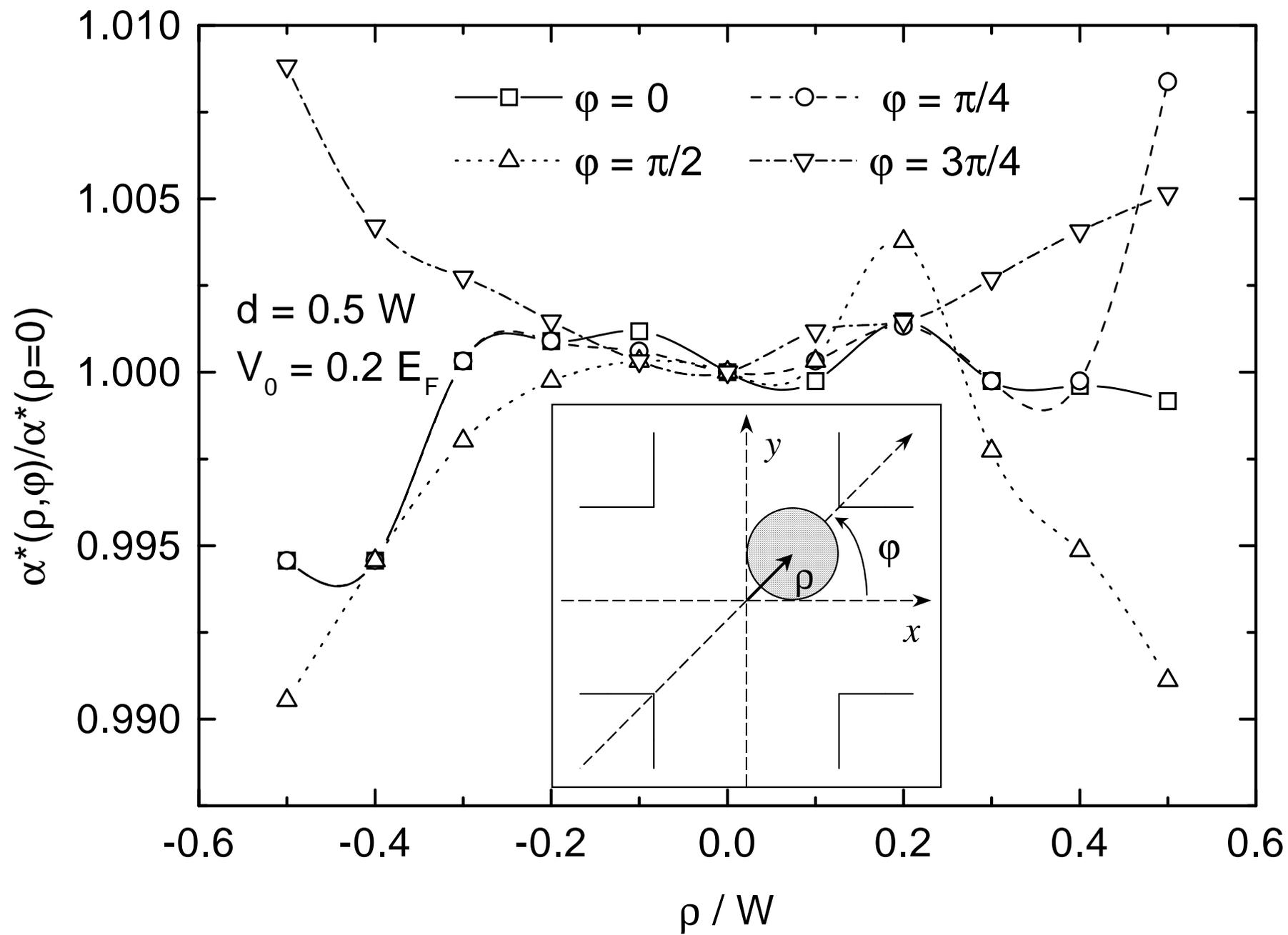
\end{document}